# BEPCII and CEPC

J. Gao, Institute of High Energy Physics,
University of Chinese Academy of Sciences, Beijing, China

*Abstract*

In this paper BEPC, BEPCII and BEPCII-U have been briefly reviewed, which lay a good foundation of CEPC as a Higgs factory. BEPCII reached the designed luminosity goal of $10^{33}$cm$^{-2}$s$^{-1}$ @1.89GeV on April 5, 2016. As an upgrade program of BEPCII, BEPCII-U has been completed and started commissioning in March of 2025 with the luminosity goal as $3.7\times10^{32}$cm$^{-2}$s$^{-1}$ @2.8GeV. CEPC as Higgs factory has been reviewed on Engineering Design Report (EDR) status and plan for construction proposal to Chinese government in 2025.

## INTRODUCTION

More than twenty years after the invention of circular collider [1], BEPC is the first electron positron collider in China completed in 1988 with the luminosity of $10^{31}$cm$^{-2}$s$^{-1}$ @1.89GeV achieved for high energy physics studies at IHEP (single ring, single bunch, single detector and normal conducting RF cavities), Precise measurement of tau lepton mass was one of the most important scientific achievements of the BEPC. The first director of BEPC project is Prof. Jialin Xie of IHEP, who received the National Scientific and Technology Progress First Prize in 2016, the highest prize in China for science and technology. The IPAC (Asia) highest prize is named with Jialin Xie. In terms of international collaboration with USA, especially with SLAC, Prof. W. Panofsky of SLAC received the International Science and Technology Cooperation Award of China in 2000 thanks to his collaborating contributions to BEPC.

To increase significantly the luminosity of BEPC, a double ring scheme has been adopted by BEPC II（multi-bunch, single detector, and superconducting rf cavities）, which was started in 2006 and completed in 2009, with designed luminosity of $10^{33}$cm$^{-2}$s$^{-1}$ @1.89GeV reached on April 5, 2016, hundred time than that of BEPC. The key path for this hundred times luminosity increase is due to following factors: the $\beta_y$ has been reduced to 1.35cm, the bunch length $\sigma_z$ has been compressed down to 1.35cm with smaller $\alpha_p$ of 0.018, the number has been increased to 118, single bunch current of 7.8mA, transverse coupling coefficient of 1%, maximum vertical beam-beam tune shift of 0.04, which agrees with the analytical estimation result also [2]. Concerning the international collaboration with KEK, Japan, Prof. Shin-Ichi Kurokawa of KEK received the International Science and Technology Cooperation Award of China in 2012 thanks to his collaborating contributions to accelerator sciences in China including BEPCII.

To extend the physics reaches of BEPCII, an upgrade program of BEPCII, so-called BEPCII-U, was formally approved in July 2021, with the luminosity goals as $1.1\times10^{33}$cm$^{-2}$s-1 @ 2.35GeV and $3.7\times10^{32}$cm$^{-2}$s$^{-1}$ @ 2.8GeV, respectively. BEPCII-U has been put to commissioning in March 2025 with some major accelerator upgrades compared with BEPCII, i.e., BEPCII-U optics has been further optimized with smaller emittance of 122nm; SC quadrupole has been replaced with a new one fabricated at IHEP with magnetic gradient of 25.2T/m, diameter of 190mm, field harmonics < $3.2\times10^{-4}$; and two addition 500MHz SC cavities compared with BEPCII (each ring has now two SC cavities in BEPCII-U of total rf voltage of 3.3MeV to compensate the synchrotron radiation energies at 2.8GeV). The total budget of BEPCII-U is 160M RBM including 3M RMB for BESIII upgrade. The BEPCII-U has been completed and started commissioning in March of 2025.

As for high energy physics development in IHEP beyond the BEPC, thanks to the discovery of Higgs boson on July 4, 2012 on LHC at CERN, in Sept. 2012, Chinese scientists proposed Circular Electron Positron Collider (CEPC) as a Higgs factory followed by a Super proton-proton Collider (SppC), both machines are situated in the same tunnel side by side [3]. The CEPC Conceptual Design Report (CDR) [4] has been completed in Nov. 2018, and accelerator Technical Design Report (TDR) has been formally released in Dec. 2023 [5]. The CEPC TDR cost has been estimated as 36.4B RMB, or about 5.2B USD. For the baseline design CEPC with 30MMW synchrotron radiation power per beam, the total AC power at Higgs energy is about 261MW.

Started from 2024, CEPC has entered Engineering Design Report (EDR) phase before construction, the geological feasibility investigations and the corresponding civil engineering design are undergoing on CEPC EDR site. CEPC detector Reference TDR design report will be completed by June 2025, and CEPC proposal will be submitted to Chinese government for the approval of construction within China's "15th five-year plan" (2026-2030). The planned construction starting time is around 2027 and the completion time around 2035 followed by 2 years commissioning before physics experiments. The operation scenario is "10-2-1-5", i.e., 10 years Higgs energy operation, 2 years Z-pole operation, 1 year of W energy operation and 5 years ttbar energy operation as an upgrade program. After the CEPC physics running, the SppC will be installed and put to operation around 2060's, with e-p and e-ion collisions possibility maintained. As for the nature of CEPC/SppC, it is an international project hosted in China with international collaborations and participations. The CEPC is developed with the aim of an early Higgs factory in the world for the HEP community worldwide.

## BEPCII AND BEPCII-U

The parameters of BEPCII and BEPCII-U are shown in Table 1, and the luminosity evolution of BEPCII is illustrated in Fig. 1 with the reached design luminosity of $10^{33}$cm$^{-2}$s$^{-1}$ @1.89GeV for physics routine operation.

The extended physics reaches of BEPCII-U at higher beam energy up to 2.8GeV is sown in Fig. 2. The comparison of the luminosities of BEPCII and BEPCII-U at higher energy are shown in Fig. 3. The main features of BEPCII and BEPCII-U are double rings with half crossing angle of 11mrad, equipped with SC focusing quadrupoles, SC rf cavities, and an S-band electron positron injection linac. The IHEP made SC 500MHz cryomodules, superconducting quadrupole, and MDI of BEPCII-U are illustrated in Fig. 4.

Table 1: The BEPCII and BEPCII-U parameters.

| Parameters | BEPCII@ 1.89GeV | BEPCII@ 2.35GeV | BEPCII-U @ 2.35GeV | BEPCII-U @ 2.8GeV |
|---|---|---|---|---|
| L [$10^{32}$cm$^{-2}$s$^{-1}$] | 10 | 3.5 | 11 | 3.7 |
| $\beta_y^*$ [cm] | 1.35 | 1.5 | 1.35 | 3.0 |
| Beam current [mA] | 920 | 400 | 900 | 450 |
| Bunch number | 120 | 120 | 120 | 120 |
| SR Power [kW] | 110 | 110 | 250 | 250 |
| Damping times [ms] | 24.7/24.7/12.3 | - | 12.8/12.8/6.4 | - |
| Half crossing angle [mrad] | 11 | 11 | 11 | 11 |
| Piwinski angle | 0.47 | - | 0.33 | - |
| $\beta_x$[m]/$\beta_y$[m] | 1/0.0135 | 1/0.0135 | 1/0.0135 | 1/0.0135 |
| $\xi_{y,lum}$ | 0.029/0.041 | - | 0.022/0.033 | 0.043 |
| Emittance [nmrad] | 122 | 147 | 152 | 200 |
| Coupling [%] | 1 | 0.53 | 0.35 | 0.5 |
| Bucket Height | 0.008 | 0.0069 | 0.011 | 0.009 |
| $\sigma_{z,0}$ [cm] | 1.15 | 1.54 | 1.07 | 1.4 |
| $\sigma_z$ [cm] | 1.35 | 1.69 | 1.22 | 1.6 |
| RF Voltage [MV] | 1.6 | 1.6 | 3.3 | 3.3 |

Over more than 40 years' design, construction and operation experiences on BEPC, BEPCII and BEPCII-U, by reaching the designed energies and luminosities with advanced technologies and large quantity of influential scientific research results obtained, IHEP and Chinese high energy physics community together with international collaborators have laid a solid ground for a next generation high energy Circular Electron Positron Collider (CEPC) after BEPCII (U).

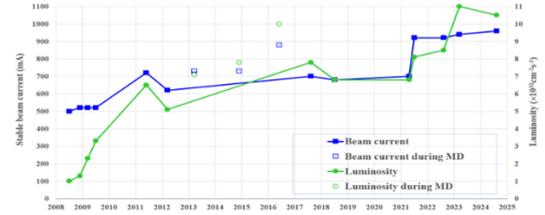

Figure 1: BEPCII luminosity evolution @1.89GeV.

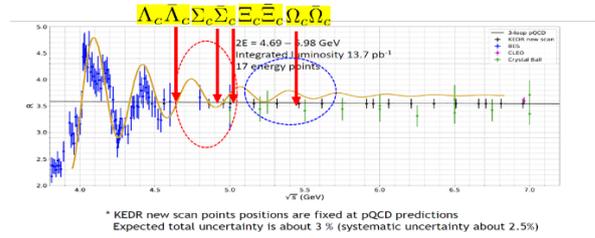

Figure 2: The physics reach of BEPCII-U.

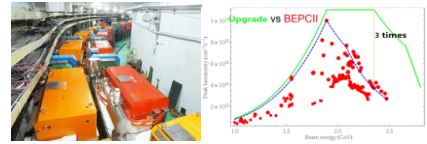

Figure 3: The peak luminosity of BEPCII vs BEPCII-U.

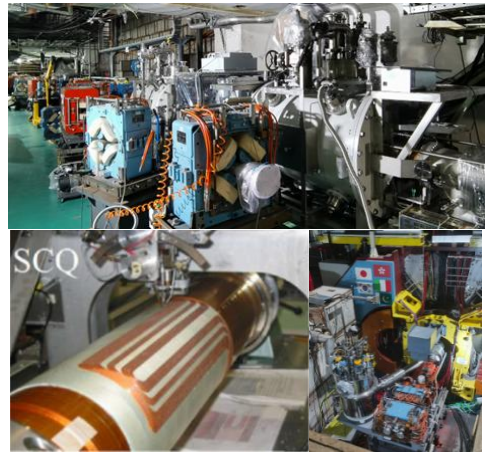

Figure 4: The IHEP made SC 500MHz cryomodules, SC quadrupole, and MDI of BEPCII-U

## CEPC ACCELERATOR

CEPC is a Higgs factory of 100km circumference composed of a 30GeV electron positron injection linac, a 1.1GeV positron damping ring, a full energy booster and a double ring collider equipped with two detectors as shown in Fig. 5, where SppC in the same tunnel of CEPC and its injection complex are also illustrated. CEPC could operate at Higg, W and Z-pole and ttbar (upgrade) energies. The physics potentials of CEPC are shown in Table 2 and the detailed Higgs and flavour physics are described in Refs. 6 and 7. Oher physics reach possibilities on Electroweak, new physics beyond Standard Model and QCD will be published later.

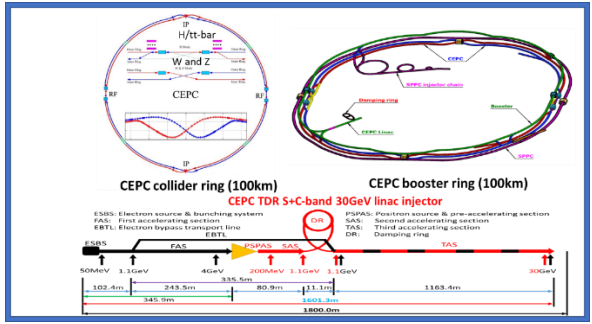

Figure 5: CEPC and SppC layout.

Table 2: CEPC physics potentials.

| Operation mode | | ZH | Z | W+W- | $t\bar{t}$ |
|---|---|---|---|---|---|
| $\sqrt{s}$ [GeV] | | ~240 | ~91 | ~160 | ~360 |
| Run Time [years] | | 10 | 2 | 1 | 5 |
| 30 MW | L / IP [×$10^{34}$ cm$^{-2}$s$^{-1}$] | 5.0 | 115 | 16 | 0.5 |
| | ∫ L dt [ab$^{-1}$, 2 IPs] | 13 | 60 | 4.2 | 0.65 |
| | Event yields [2 IPs] | 2.6×$10^6$ | 2.5×$10^{12}$ | 1.3×$10^8$ | 4×$10^5$ |
| 50 MW | L / IP [×$10^{34}$ cm$^{-2}$s$^{-1}$] | 8.3 | 192 | 26.7 | 0.8 |
| | ∫ L dt [ab$^{-1}$, 2 IPs] | 21.6 | 100 | 6.9 | 1 |
| | Event yields [2 IPs] | 4.3×$10^6$ | 4.1×$10^{12}$ | 2.1×$10^8$ | 6×$10^5$ |

Table 3: CEPC collider parameter.

| | Higgs | Z | W | $t\bar{t}$ |
|---|---|---|---|---|
| Number of Ips | \multicolumn{4}{c}{2} | | | |
| Circumference (km) | 99.955 | | | |
| SR power per beam (MW) | 30 | | | |
| Energy (GeV) | 120 | 45.5 | 80 | 180 |
| Bunch number | 268 | 11934 | 1297 | 35 |
| Emittance $\varepsilon_x/\varepsilon_y$ (nm/pm) | 0.64/1.3 | 0.27/1.4 | 0.87/1.7 | 1.4/4.7 |
| Beam size at IP $s_x/s_y$ (um/nm) | 14/36 | 6/35 | 13/42 | 39/113 |
| Bunch length (natural/total) (mm) | 2.3/4.1 | 2.5/8.7 | 2.5/4.9 | 2.2/2.9 |
| Beam-beam parameters $x_x/x_y$ | 0.015/0.11 | 0.004/0.127 | 0.012/0.113 | 0.071/0.1 |
| RF frequency (MHz) | 650 | | | |
| Luminosity per IP ($10^{34}$ cm$^{-2}$ s$^{-1}$) | 5.0 | 115 | 16 | 0.5 |

The CEPC collider EDR baseline parameters are shown in Table 3. CEPC is a crab-waist collision collider, the maximum luminosities at different energies could be also estimated by J. Gao's analytical formulae [2][8] including crab waist effects. The general maximum luminosity formula of a lepton collider per IP is shown in Eq. 1 [8] as follows:

$$L_{max}[cm^{-2}s^{-2}] = 0.158 \times 10^{34} \frac{(1+r)}{\beta_y[mm]} I_b[mA] \times \sqrt{\frac{U_0[GeV]}{N_{IP}}} e^{\frac{\sqrt{\Phi_p}}{3.22}} (1 + 0.000505\Phi_p^2). \quad (1)$$

where $r=\sigma_y/\sigma_x$ and $\beta_y$ are the values at interaction point (IP), $U_0$ is the synchrotron radiation energy loss per tune of a charged particle, $N_{IP}$ is the number of IP, $I_b$ is the beam current, and $\Phi_p$ is the Piwinski angle. For a collider of isomagnetic lattice like CEPC, by using $U_0=C_\gamma E^4/R$ and the beam power $P_b=I_bU_0$, Eq. (1) could be expressed also in Eq. (2) as follows:

$$L_{max}[cm^{-2}s^{-2}] = 0.158 \times 10^{34} \frac{(1+r)}{\beta_y[mm]} \sqrt{\frac{R[m]}{C_\gamma[mGeV^3]N_{IP}}} \times \left(\frac{P_b[MW]}{E[GeV]^2}\right) e^{\frac{\sqrt{\Phi_p}}{3.22}} (1 + 0.000505\Phi_p^2). \quad (2)$$

where $C_\gamma=8.85\times10^{-5}$mGeV$^{-3}$, R is the local bending radius of dipoles of the collider ring and E is the beam energy. By using the CEPC collider ring parameters in Table 3 and Eq. 2, one could obtain the CEPC maximum luminosities per IP ($10^{34}$ cm$^{-2}$ s$^{-1}$) at Higgs, W, Z-pole and ttbar energies as 5, 115, 12, 0.59, respectively, which are very close to the luminosities shown in Table 3 obtained by numerical simulations taking into account of crab-waist collision effects [5]. For a collider of non-isomagnetic nature, like Super KEK B, Eq. (1) could be used to estimate the maximum luminosity.

As for a hadron collider, such as SppC, FCC$_{hh}$ or LHC, the maximum beam-beam tune shift value per IP could also be estimated by J. Gao's analytical formulae [2][8] both for round and flat beams, as expressed in Eq. (3) [8]:

$$\xi_{max,y} = \frac{H_0 \gamma}{f(x_*)} \left(\frac{r_p}{6\pi R N_{IP}}\right)^{1/2} F. \quad (3)$$

where $H_0$=2845, $\gamma$ is the normalized beam energy, $r_p$ is the hadron (proton) radius, R is the local dipole bending radius, $N_{IP}$ is the number of interaction point, F=(4√2)/3 for round beam collision, F=1 for flat beam collision, f(x) is expressed in Eq. 4, and $x_*$ can be solved by Eq. (5).

$$f(x) = 1 - \frac{2}{\sqrt{2\pi}} \int_0^x e^{-t^2/2} dt. \quad (4)$$

$$x_*^2 = \frac{4f(x_*)^2}{H_0 \pi \gamma} \left(\frac{6\pi R}{r_p N_{IP}}\right)^{1/2}. \quad (5)$$

By using the parameters of SppC [5] and Eqs. (3)-(5) for round beam collision, the analytically estimated $\xi_{max,y}$ is 0.0147, and the numerical estimation value of $\xi_{max,y}$ for SppC is 0.015. As for LHC, the analytical estimated $\xi_{max,y}$ value is 0.0045 by using Eqs. (3)-(5) [8], and the $\xi_{max,y}$ value in the LHC parameter table is 0.005, and the LHC experimental result is about 0.0045 [9].

Concerning CEPC accelerator technology development in TDR and EDR phases, a full spectrum accelerator key technology have been developed such as the 1.3GHz and 650MHz SC accelerator cryomodules shown in Figs. 6 and 7, high efficiency CW 650MHz klystron and 80MW high power C-band klystron as shown in Figs. 8 and 9, automatic fabrication lines for booster dipole magnets and

NEG coated vacuum chambers as shown in Fig. 10, and the mockup tunnel as shown in Fig. 11. All CEPC construction related industrial mass production preparations will be completed within EDR phase from 2024 to 2027, and CEPC is expected to start the construction around 2027.

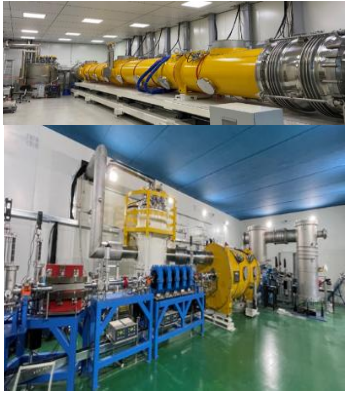

Figure 6: The CEPC 1.3GHz cryomodule horizontal test reached design goal with $E_{acc}$ =23.1MV/m and $Q_0$= 3.4$\times$ $10^{10}$ (up); A short 650MHz cryomodule has been completed in TDR (down).

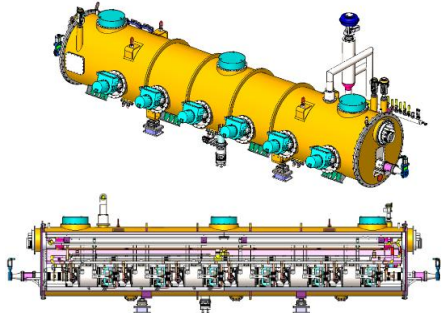

Figure 7: The CEPC 650MHz full size cryomodule for collider ring.

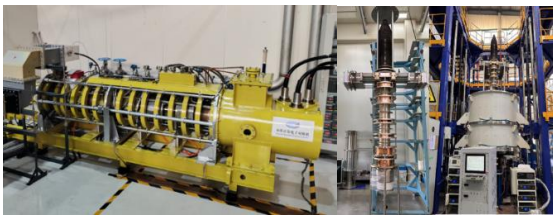

Figure 8: The CEPC 650MHz 803kW CW klystron with efficiency 78.5% achieved in 2024 (left); CEPC multibeam 650MHz 800kW CW klystron with efficiency 80.3% (right).

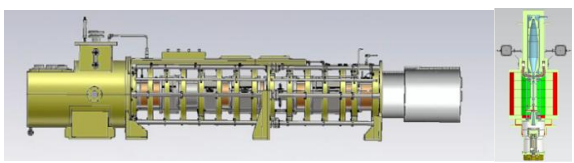

Figure 9: The CEPC 650MHz one stage energy recovery klystron with efficiency of 85% (left); CEPC C-band klystron of 5720 MHz, 80MW, 3us, 100Hz (right).

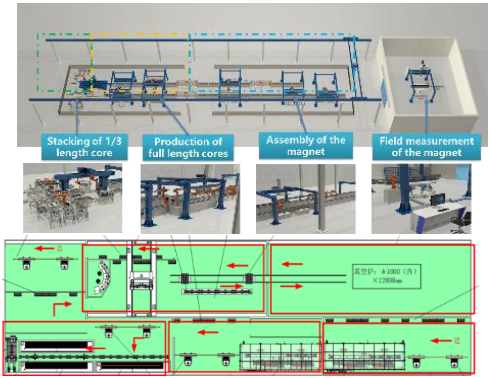

Figure 10: The CEPC automatic fabrication lines for booster dipole magnets (up) and vacuum chamber NEG coating (down).

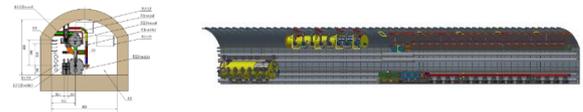

Figure 11: The CEPC 60m long mockup tunnel.

## CEPC DETECTOR

The CEPC detector reference TDR design is shown in Fig. 12, and the baseline detector technologies and performances have been identified specified as shown in Table 4.

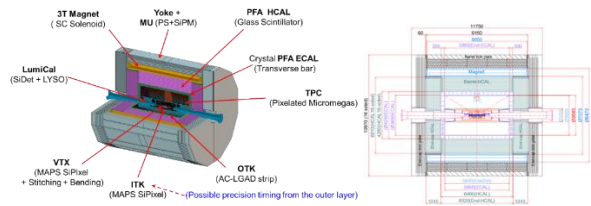

Figure: 12 The CEPC detector in reference TDR design.

The subdetector technologies have been developed, such as Vertex Detector in collaborative R&D with the ALICE team on stitching CMOS technology; Silicon Trackers including Inner Tracker (HV-CMOS) and Outer Tracker (AC-LGAD strip); TPC mechanical and cooling design completed with the prototype pixelated micromegas module ready for test beam; Electromagnetic Calorimeter long crystal bar developed; Glass Scintillator HCAL mass production of full size (4$\times$4$\times$1 cm$^3$) GS samples demonstrated; Muon Detector with 4m long strip plastic scintillators produced; Superconducting Magnet with aluminum stabilized $Nb_bT_i$ Rutherford Cable developed; MDI design and more sub-systems have been developed, such as electronics, TDAQ, software/computing, mechanical integration, performance and detector cost estimations, etc. CEPC detector technology development has wide international collaborations and has participated in the ECFA DRD program in depth as shown in Table 5.

In addition to the CEPC detector design, technology development and integration, the CEPC detector and auxiliary halls have also been designed in a preliminary way as shown in Fig. 13 together with CEPC accelerator civil engineering designs. With the above illustrated progresses, the CEPC detector reference TDR has been reviewed by CEPC Detector Review Committee (IDRC) on April 14-16, 2025, and CEPC detector reference TDR will be released by mid of 2025 [10].

Table 4: The CEPC reference TDR detector technologies and specifications.

| Sub-system | Key technology | Key Specifications |
|---|---|---|
| Vertex | 6-layer CMOS SPD | $s_{rf} \sim 3$ mm, $X/X_0 < 0.15\%$ per layer |
| Tracking | CMOS SPD ITK, AC-LGAD SSD OTK, TPC + Vertex detector | $\sigma\left(\frac{1}{P_T}\right) \sim 2 \times 10^{-5}$ $\oplus \frac{1 \times 10^{-3}}{P \times \sin^{3/2}\theta}(GeV^{-1})$ |
| Particle ID | dN/dx measurements by TPC, Time of flight by AC-LGAD SSD | Relative uncertainty ~ 3% $s(t) \sim 30$ ps |
| EM calorimeter | High granularity crystal bar PFA calorimeter | EM resolution ~ $3\%/\sqrt{E(GeV)}$ Effective granularity ~ $1 \times 1 \times 2$ cm$^3$ |
| Hadron calorimeter | Scintillation glass PFA hadron calorimeter | Support PFA jet reconstruction Single hadron $\sigma_E^{had} \sim 40\%/\sqrt{E(GeV)}$ Jet $\sigma_E^{jet} \sim 30\%/\sqrt{E(GeV)}$ |

The CEPC detector technology development has wide international collaborations and has participated in the ECFA DRD program as shown in Table 5.

Table 5: CEPC detector in the ECFA DRD program.

| Sub-system | DRD | Sub-system | DRD | Sub-system | DRD |
|---|---|---|---|---|---|
| Pixel Vertex Detector | 3 | Electromagnetic Calorimeter | 6 | Super Conducting Magnet | |
| Inner Silicon Tracker | 3 | Hadron Calorimeter | 4, 6 | Mechanical and Integration | 8 |
| Outer Silicon Tracker | 3 | Machine Detector Interface | 8 | General Electronics | (7) |
| Gas Tracker (TPC / DC) | 1 | Luminosity Calorimeter | | Trigger and DAQ | (7) |
| Muon Detector | 1 (RPC) | Fast Luminosity Monitor | 3 | Offline Software | |

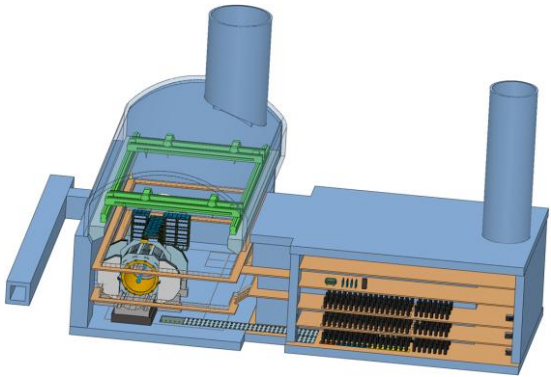

Figure 13: The CEPC detector experimental and auxiliary halls in reference TDR design.

## IHEP AS CEPC HOST LAB

The CEPC as the largest future collider in the world is very challenging in its design, R&D, construction, operation, experiments and international collaborations, and the requirement for its host lab is also very demanding. IHEP as the CEPC host lab has distinguished features spanning large range of scientific research areas, such as BEPCII (U) in IHEP Beijing Campus for collider based particle physics studies; neutrino physics experiments, such as Daya Bay reactor Neutrino Experiment (retired), Jiangmen Underground Neutrino Observatory JUNO, High-energy Underwater Neutrino Telescope (HUNT) underwater in south China Sea (planned); Yang Ba Jing International Cosmic Ray Observatory (retired), Large High-Altitude Air Shower Observatory (LHAASO); Ali CMB Polarization Telescope (AliCPT); Gravitational wave EM Counterpart All-sky Monitor (GECAM), Insight Hard X-ray Modulation Telescope (HXMT), the High Energy cosmic-Radiation Detection (HERD) to be placed on Chinese Space Station in 2027; 4th generation High Energy Photon Source (HEPS) in IHEP Beijing Huairou Campus, Chinese Spallation Neutron Source (CSNS I and II) in IHEP Dongguan Campus; Nuclear Technology Application Center and Advanced Mechanical Fabrication Center in IHEP Jinan Campus. In addition to domestic based experiments, IHEP has participated wide range international collaborations, such as with CERN on LEP, LHC, and HL-LHC; with KEK on Belle-II of Super KEK B, and ILC, etc. There are over 1500 full-time staff, as well as over 1000 postdocs and graduate students working and studying at IHEP.

Concerning the CEPC civil construction, the previous large civil engineering experiences gained on IHEP's non-accelerator based projects in Daya Bay and JUNO neutrino experiments are very useful to the construction of CEPC's huge deep underground experimental halls, for example, the JUNO detector hall, with dimensions of 56.25m×49m ×27m, has similar sizes as that of CEPC detector experimental halls, as shown in Fig. 14.

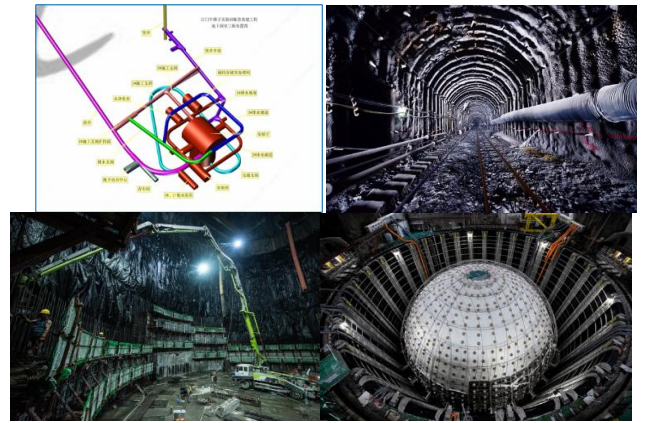

Figure 14: The JUNO civil engineering complex and detector experimental hall.

## CEPC COLLABORATIONS

Concerning CEPC domestics and international collaboration, CEPC has been developed with the guidance and helps from Institutional Board (IB), International Advisory Committee (IAC), International Accelerator Review Committee (IARC), International Accelerator Review Committee (IDRC), and CEPC accelerator TDR (+cost) review committees. Both CEPC CDR and TDR have more than one thousand authors from around world of more than hundreds of international institutes and universities and several tenths of countries. Since 2014, every year CEPC international workshop is held in China, since 2018, each year, CEPC workshop of European Edition or US edition is held, and since 2015, each year, CEPC participates HKUST IAS HEP conference and relevant mini-workshops, with participants from worldwide. Till now, more than 20 MoUs have been signed between IHEP and international institutes and Universities on CEPC collaborations. In terms of industrial collaboration and preparation for CEPC construction, in Nov. 2017 at IHEP, CEPC Industrial Promotion Consortium (CIPC) has been established, and CIPC has made important contributions to the CEPC's CDR, TDR and EDR developments.

Aside from the preparation of CEPC proposal to Chinese government in 2025 for the "15[th] five year plan" with CEPC accelerator TDR and detector reference TDR completed and EDR in progress including EDR site geological feasibility and civil engineering studies with strong local government supports, CEPC actively participated also to European Strategy for Particle Physics Upgrade 2026 [11] together with other Higgs factory projects, such as $FCC_{ee}$, ILC, CLIC, HALHF, etc. progressing forwards towards the common goals.

## CHINA AS CEPC HOST COUNTRY

CEPC as a Higgs factory of 100km circumference, its investment, technologies' development, experienced personnel, industrial construction capabilities, etc. are very demanding for its host country. China as CEPC host country, within last 7 years, has invested many accelerator-based projects as shown in Table 6. The total investment of the approved projects is about 39B RMB, higher than the CEPC total cost of 36.4BRMB (5.2B USD). The realization of these projects will promote significantly the relevant personnel resources, industrial fabrication and mass production capabilities. As for trying to realize a green and sustainable Higgs factory, the CEPC has been optimized with the aim of maximizing scientific outputs with minimum construction and operation cost [12] relating also to lowering $CO_2$ equivalent emissions [13], through machine design and key technology development, such as increasing the efficiency of accelerator components and investigating low temperature heat recovery technologies, etc. As for the type of energy used by CEPC, it depends greatly on the construction site, however, in general, since April 2025, China has become the largest nuclear power electricity production country in the world including existing and undergoing construction nuclear power plants, reaching $1.13 \times 10^5$MW.

Table 6: CEPC in synergy with other accelerator-based projects in China

| Project name | Machine type | Location | Cost (B RMB) | Completion time |
|---|---|---|---|---|
| **CEPC** | Higgs factory Upto ttar energy | Led by IHEP, China | 36.4 (where accelerator 19) | Around 2035 (starting time around 2027) |
| BEPCII-U | e+e-collider 2.8GeV/beam | IHEP (Beijing) | 0.15 | 2025 |
| HEPS | 4[th] generation light source of 6GeV | IHEP (Huanrou) | 5 | 2025 |
| SAPS | 4th generation light source of 3.5GeV | IHEP (Dongguan) | 3 | 2031 (in R&D, to be approved) |
| HALF | 4th generation light source of 2.2GeV | USTC (Hefei) | 2.8 | 2028 |
| SHINE | Hard XFEL of 8GeV | Shanghai-Tech Univ., SARI and SIOM of CAS (Shanghai) | 10 | 2027 |
| S3XFEL | S3XFEL of 2.5GeV | Shenzhen IASF | 11.4 | 2031 |
| DALS | FEL of 1GeV | Dalian DICP | - | (in R&D, to be approved, ) |

In 2035, the nuclear power in China will be around 10% of the total electric power produced. In terms of other types of electricity, since April 2025, the clean wind and photovoltaic electricity powers produced in China have surpassed that from the coals. Concerning dual carbon strategy, China's goal is to achieve peak $CO_2$ emissions before 2030 and carbon neutrality before 2060. The clean electricity resource development in China is favorable for CEPC construction and operation in China.

## CONCLUSIONS

In this article BEPCII(U) and CEPC have been reviewed in a general way, their inter-relations have been underlined. The readiness of both accelerator and detector technologies of CEPC through TDR and EDR, the general background for the host lab (IHEP) and host country (China), and timeline for CEPC proposal and construction timeline have been introduced. As an international project, CEPC welcome international participations and contributions worldwide.

## ACKNOWLEDGEMENTS

The author thanks BEPCII (U) and CEPC teams, the international and industrial collaborators, for their tireless efforts towards establishing excellent colliders for high energy physics experiments.